# Transparent low-density stellar plasma extending over very large volumes with large masses and Dark Haloes

**Y. Ben-Aryeh**

Technion-Israel Institute of Technology, Physics Department, Israel, Haifa 32000

E-mail: phr65yb@physics.techion.ac.il

**The free electron model with Boltzmann statistics for spherical low-density plasmas (Scientific Reports 9. 20384, 2019) is developed further by numerical calculations with asymptotic relations obtaining the density of electrons, mass densities and the potentials of such plasmas. Solutions are developed as function of a pure number $x$ proportional to the distance from the stellar plasma center (galaxy center) with extremely small coefficient so that these solutions are essentially functions of large astronomical distances and masses. The present plasma is divided into a central part and very long tail where a part of the large mass of this plasma is included in the long stellar plasma tail. The present model is specialized to completely ionized Hydrogen plasma where emission and absorption of spectral lines can be neglected in the low density stellar plasma. It is shown that the present low-density plasma might represent dark halo which permeates and surrounds the compact galactic stars. Such plasma is found to be transparent in most of the EM spectrum.**

## I. INTRODUCTION

In previous work [1] we studied the stability of low-density star plasma where the stability is produced by the balance between the gravitational forces and the pressure produced by perfect gas conditions, and found the impact of such model on the properties of low-density plasmas. While we treated this topic in the previous work [1] we obtained there only approximate analytical results for the central body of such stellar plasma. By using now numerical solutions (using Mathematica NDSolve) we get in the present work explicit numerical solutions. These numerical calxulations help us to get asymptotic equations for the change of the ionized electrons density $n_i(r)$ and the potential $\phi(r)$ as function of the distance $r$ from the star plasma center. I assume a stationary state for which Boltzmann distribution is valid. Stellar plasma behaves as a perfect gas at densities lower than a critical value. This critical value is given by the condition that the Coulomb interaction energy is smaller than the thermal energy [2]. We assume one ionic component with atomic number $Z$ but the analysis can be easily generalized if we have more ionic components.

Quantum effects for low density ionized stellar plasmas, with one ionic component, with atomic number $Z$ (satisfying Boltzmann statistics) can be neglected under the condition $Ze^2/4\pi\varepsilon_0\hbar v \gg 1$ where $\hbar$ is Planck constant (divided by $2\pi$), $e$ is the electron charge and $v$ the electron velocity [3-5]. Similar relations can be obtained if we have more than one ionic component. Under the condition: $Ze^2/4\pi\varepsilon_0\hbar v \geq 1$ we can still use a classical analysis and the quantum effects result as corrections to the classical formulas (Gaunt factors [6] which under this condition are small numbers). So the present analysis in which quantum effects are neglected is developed under the condition

$$v_{e,\max} \leq \frac{Ze^2}{4\pi\varepsilon_0} \cong 2Z \cdot 10^6 \left(\frac{m}{\sec}\right) \quad , \tag{1}$$

where the subscript $e,\max$ refers to the maximal electrons velocities for which the present Boltzmann distribution is valid.

We assume for such stellar plasma that the temperature is relatively high so that the present Boltzmann statistics is valid. For a perfect gas i.e., when the plasma ion density is sufficiently rarefied with ionized electron density $n_i(m^{-3})$, at temperature $T$, the electrons pressure $P_i$ is given by

$$P_i = n_i k_B T \quad , \tag{2}$$

where $k_B$ is Boltzmann constant, $T$ the absolute temperature and $n_i$ is the number of ionized-electrons per unit volume. The number of neutral atoms relative to ionized electrons can be calculated by Saha's equation [7, 8]. Such calculations are very complicated but for Hydrogen plasma, which is considered as the main component of the low density stellar plasma, simple analytical result can be obtained as [7]

$$\frac{n_i}{n_n} \cong 2.4 \cdot 10^{21} \frac{T^{3/2}}{n_i} \exp\left(\frac{-13.6 eV}{k_B T}\right) \quad , \tag{3}$$

where $\frac{n_i}{n_n}$ is the ratio between the number of ionized electrons $n_i(m^{-3})$ and the number $n_n(m^{-3})$ of neutral hydrogen atoms and by conversion to MKS unit $13.6 eV \cong 2.179 \cdot 10^{-18} J$. For temperature 20000(K) and ionized electron density $n_i = 10^6 (m^{-3})$ we have $4.84 \cdot 10^{18}$ ions per one neutral hydrogen atom! So under this condition absorption and emission of radiation related to spectral lines can be neglected and the low density plasma can be considered as dark matter. One should take into account that although in this calculation the thermal energy is small relative to the ionization energy the ionization is very close to unity due to the effective statistical weight of the continuum spectrum which is very high (proportional to $\left(\frac{2\pi m_e k_B T}{h^2}\right)^{3/2} = \left(\frac{1}{\lambda_{DB}}\right)^3$ where $m_e$ is the electron-mass, $h$ is Planck constant and $\lambda_{DB}$ is the De Broglie wavelength [8].) For higher temperatures and/or lower densities this ratio between ionized electrons and neutral atoms becomes even much higher. But for temperature, lower than $20000(K)$ neutral Hydrogen atoms should be taken into account, when the density of ionized electrons is enough large. One should take into account that in the same temperature for which low density plasma is transparent the high density stars can have strong absorption and emission of radiation. This fact is related to the $1/n_i$ dependence in Eq. (3) where for high density stars the density $n_i$ becomes very large reducing very much the ratio between the density of ionized electrons and the neutral Hydrogen atoms. In contrast to low density stellar plasma the high density stars can lead to emission and absorption of radiation related to spectral lines of Hydrogen and other atoms included in the stars atmospheres.

One should also notice that for longer radial distances from the stellar plasma center (where the density of ionized electrons is decreasing), according to Eq. (3) the ratio between ionized electrons density $n_i$ to the neutral Hydrogen atoms density $n_n$ is increasing inversely proportional to $n_i$. So, this ratio is increasing further so that the approximation in which we neglect emission and absorption of spectral lines for low density plasma is even much better.

Taking into account composition of the stellar plasma with different atoms and different temperatures we use for simplicity of discussion the following semi-empirical relation

$$\rho(r) = n_i(r) \kappa m_N \quad , \tag{4}$$

where $n_i(r)(m^{-3})$ and $\rho(r)(kg/m^3)$ are, respectively, the ionized electrons density and the mass density given as function of the distance $r$ from the stellar plasma center, $\kappa$ is the number of nucleons per ionized electron, and $m_N = 1.67 \cdot 10^{-27} \, kg$ is the nucleon mass. For completely ionized Hydrogen plasma $\kappa = 1$ but for partial ionized plasma $\kappa$ is larger. For a certain composition of the stellar plasma and certain temperature we may assume $\kappa$ to be a certain empirical parameter. We introduce the parameter $\kappa$ for generality of the analysis but as we specialize later the present analysis to completely ionized Hydrogen plasma we assume there $\kappa = 1$.

There are various models describing dark matter haloes [9-14]. While such models give a good agreement with astronomical observations such as gravitational lensing [15] the composition and the physical nature of such dark matter Haloes is not clear. In the previous work [1] I related the dark Haloes to low density ionized transparent plasma. The absorption and emission of such plasma is proportional to multiplication of the electrons and ions extremely low densities so that such effects can be neglected. But, as such low density plasma extends over extremely large volume gravitational effects become important.

In the present work we would like to generalize our previous work so that the total potential produced by the galaxy is composed by the superposition of the potential of the compact high density stars with that of the low density stellar plasma. While the luminosity of the galaxy stars follows Hertzsprung-Russel (HR) diagrams [16, 17] giving the stars luminosity as function of temperature, the emission and absorption of radiation by the low density plasma depends on the emissivity which is the ratio between the radiation intensity emitted by a certain body and that of black body radiation. For any material which is in thermal equilibrium the emissivity should be equal to the absorptivity. For very low density plasma in which the absorption and emission of radiation vanish, the emissivity tends also to zero and HR diagrams do not apply for such system. Following these ideas we would like to discuss in the present work the possibility that superposition of the high density stars potentials with that of low density stellar plasmas would explain the density profiles of dark haloes.

I suggested in the previous article [1] that the present analysis describes some kind of dark halo. I would like to check this idea by extending the previous treatment of the electromagnetic (EM) properties of low-density plasma to that of the general free electrons EM model [18]. This model depends on two parameters: 1) Plasma frequency $\omega_p$ which is given by

$$\omega_p^2 = \frac{n_e e^2}{\varepsilon_0 m_e} \cong 3.18 \cdot 10^3 n_e \tag{5}$$

where $n_e$ ($m^{-3}$) is the ionized electrons density and $m_e$ is the electron mass. For very low ionized electrons densities $\omega_p$ is in the radio frequency region. 2) Important parameter in this model is $\gamma$ representing the decay constant. In the previous work [1] thermal bremsstrahlung effects were treated which are influencing the absorption and emission of EM radiation by low density plasma. Such radiative processes are proportional to the products of the electrons and ions densities so that they become very small for very low-density plasmas, and consequently the decay constant $\gamma$ becomes extremely small.

Near and above the plasma frequency which is in the far radio frequency region, dispersion effects might be observed. Gravitational lensing which is achromatic leads to bending of the EM beam from the low density of the star to its high density, so, it behaves like converging lens for massive opaque stars [15]. Here for low density plasma above and near the plasma frequency the bending of the EM beam is from the high density to the low density, and it is opposite to gravitational lensing. So, in this frequency (in the far radio frequencies near the plasma frequency) the plasma might behave as diverging lens. Divergence due to plasma has been related to extra galactic sources and also was related to refractive radio phenomena [19, 20]. Due to in-homogeneity in the plasma density I expect broadening effects in the plasma frequency but I hope that it might be possible to observe such dispersion for low density star plasmas at frequencies above and near the plasma frequencies, in the far radio frequencies region. Such observation if it could be obtained in dark matter would lead to clear proof for the relevance of low-density plasma to dark matter.

## 2. THE STABILITY OF LOW DENSITY STELLAR PLASMA ANALYZED BY THE USE OF BOLTZMANN STATISTICS WITH ASYMPTOTIC RELATIONS

For simplicity of discussion we treat low density stellar plasma with spherical symmetry where its stability is related to gravitational forces. We simplify the treatment by assuming one ionic component with atomic number $Z$ but the analysis can easily be generalized if we have more ionic components.

The gravitational force $g$ at a distance $r$ from the spherical stellar plasma center is due entirely to the mass $M_r$ interior to this distance:

$$g = -GM_r / r^2 \quad , \tag{6}$$

where $G$ is constant of gravitation. We denote $r$ and $\phi$ as the variable distance and the potential of the low-density plasma from its center, respectively. Assuming that for the stellar plasma, gravitational potential $\phi$ has spherical symmetry then

$$g = -d\phi(r)/dr \quad , \tag{7}$$

According to the hydrostatic equation for isothermal free electron model

$$dP = g\rho dr \qquad . \tag{8}$$

where $P$ is the stellar plasma pressure and $\rho$ the stellar-mass density, both are functions of distance $r$ from the stellar plasma center (galaxy center). This equation describes the decrease of the plasma pressure (proportional to decrease of the ionized electron density) as we move to larger values of $r$ opposing the attractive gravitational forces.

Assuming that the stellar plasma behaves as an ideal gas then the pressure $P$ is given by Eq. (2) as $P = n_i k_B T$. Assuming also that the gradient of temperature is small relative to the gradient of the electron density $n_e$ (isothermal process) then we get:

$$dP = k_B T dn_i = -\rho d\phi \qquad . \tag{9}$$

Substituting $\rho$ from Eq. (4) into Eq. (9) we get

$$k_B T dn_i(r) = -\kappa m_N n_i(r) d\phi(r) \rightarrow \frac{dn_i(r)}{n_i(r)} = -\frac{\kappa m_N}{k_B T} n_i \qquad . \tag{10}$$

In the present free electrons isothermal model we assume that $\kappa$ is representing the number of nucleons per one ionized electrons which is a certain averaged constant.

This equation has the general solution

$$n_e(r) = n_0 \exp\left[-\left(\frac{\kappa m_N}{k_B T}\right)\phi(r)\right] \quad ; \quad \phi(r=0) = 0 \quad ; \quad n_i(r=0) = n_0 \, , \tag{11}$$

where $n_0 (m^{-3})$ is the ionized electrons density in the center of the stellar plasma (taken as experimental parameter) and by our definition the potential $\phi$ vanish at the stellar plasma center.

One should notice that the potential $\phi(r)$ is given in unit of $\left(\dfrac{k_B T}{\kappa m_N}\right)$ i.e. unit of velocity squared ( $m^2/\sec^2$ ) as the exponent in Eq. (1) is a pure number. Using stability-conditions the equation for the potential $\phi(r)$ is given as [1]:

$$\frac{d^2\phi(r)}{dr^2} + \frac{2}{r}\frac{d\phi(r)}{dr} = 4\pi G \rho = 4\pi G \kappa m_N n_i = 4\pi G \kappa m_N n_0 Exp\left[-\left(\frac{\kappa m_N}{k_B T}\right)\phi(r)\right]. \quad (12)$$

Here $G$ is the gravitational constant and we inserted here the relation $\rho(r) = n_i(r)\kappa m_N$ where $n_i(r)$ is given by Eq. (11).

Substituting

$$x = r\sqrt{\xi} \quad ; \quad \xi = 4\pi G \frac{(\kappa m_N)^2 n_0}{k_B T} = 1.69 \cdot 10^{-40} \frac{n_0 \kappa^2}{T} \quad (m^{-2}) \quad (13)$$

where $\xi$ is extremely small number (under the present conditions), Eq. (12) is changed to

$$\frac{d^2\phi(x)}{dx^2} + \frac{2}{x}\frac{d\phi(x)}{dx} = \left(\frac{k_B T}{\kappa m_N}\right) Exp\left[-\left(\frac{\kappa m_N}{k_b T}\right)\phi(x)\right] = \frac{1}{b} Exp\left[-b\phi(x)\right] \quad ; \quad b = \frac{\kappa m_N}{k_b T}. \quad (14\text{-a})$$

One should take into account that $x$ is a normalized pure number where for $x = 1$ the distance $r$ from the star center stretches to extremely very long distance equal to: $1/\sqrt{\xi}$ and where according to Eq. (13) $\sqrt{\xi}$ is a very small number. Our solutions for the number of ionized electrons and the potentials will be given as function of the pure numerical parameter $x$ i.e. as $n_e(x)$, and $\phi(x)$. So, by transforming our solutions to those of functions of the distance $r$ from the stellar plasma center they become functions of astronomical long distances. One should notice that Eq. (14-a) is given as function of the parameter $b = \frac{\kappa m_N}{k_b T}$ where $\kappa$ and $T$ are taken as experimental constants, but for Hydrogen completely ionized plasma $\kappa = 1$.

Using numerical calculations by Mathematica Eq. (14-a) was solved for different $b$ values. Then, by substituting the numerical $\phi(r)$ values in Eq. (11) the ionized electron density $n_i(r)$ can be calculated. This is done by using the following transformation for $\phi(x)$ of Eq. (14-a):

$$\phi(x) \Rightarrow \phi\left(R\sqrt{\xi}\right) \quad ; \quad \xi = 1.69 \cdot 10^{-40} \frac{n_0 \kappa^2}{T} \quad (m^{-2}) \quad . \quad (14\text{-b})$$

In the following analysis we would like, however, to develop asymptotic equations to Eqs. (14-a), and (14-b), checking their validity by the numerical calculations.

In the previous work [1], by using Eq. (11), we performed the first and second derivatives of $n_i(r)$ according to $r$ and related such derivatives to the corresponding derivatives of $\phi(r)$. By substituting these relations in Eq. (12) and using the relation $x = r\sqrt{\xi}$ we obtained the differential equation for $\theta(x) = \dfrac{n_i(x)}{n_0}$ which is given as [1]:

$$\frac{\partial^2 \theta(x)}{\partial x^2} - \frac{2}{x}\frac{\partial \theta(x)}{\partial x} + \frac{1}{\theta(x)}\left(\frac{\partial \theta(x)}{\partial x}\right)^2 = \theta(x)^2 \quad ; \quad x = r\sqrt{\xi} \quad ; \quad \theta(x) = \frac{n_i(x)}{n_0} \quad . \quad (15)$$

It is quite easy to find that Eq. (15) is satisfied for

$$\theta(x) = \frac{n_i(x)}{n_0} = \frac{2}{x^2} \rightarrow n_i(x) = n_0 \frac{2}{x^2} \quad ; \quad n_i(r) = n_0 \frac{2}{\xi r^2} \quad . \quad (16)$$

Here the parameter $x = r\sqrt{\xi} \gg 1$ where $r(m)$ is the distance from the stellar-plasma center (galaxy center) and $\xi(m^{-2})$ is a very small number. But as the solution of Eq. (15) by Eq. (16) does not satisfy the boundary condition $\theta(x=0) = \dfrac{n_i(x=0=r)}{n_0} = 1$ this solution is valid only for $x \gg 1$. We will use the asymptotic relation of Eq. (15) to get an approximate analytical equation for the stellar plasma ionized electrons density as function of the distance from the stellar plasma center (equal to the distance from the galaxy center) for cases for which $x \gg 1$.

By numerical calculations (using Mathematica NDSolve) we obtained exact numerical solutions of Eq. (15). Such solutions give the density of electrons $n_i(x)\,(m^{-3})$ as function of the distance $r = \dfrac{x}{\sqrt{\xi}}\,(m)$ from the star center, and as function of the density of electrons $n_o\,(m^{-3})$ in the star center, which is taken as experimental parameter. These solutions enable us to get asymptotic results at large distance $r$ from the star center, for the density of ionized electrons. We find that in addition to the central part of such stellar plasma (treated analytically in our previous work [1]) we get now a very long tail of the stellar plasma potential and its density

An approximate solution is obtained for the region near the center of the star by using series expansion of the exponential function of Eq. (11). Then, we get

$$\theta(x) = \frac{n_e(x)}{n_0} = Exp\left[-\frac{1}{8}x^2\right] \rightarrow \frac{n_e(r)}{n_0} = Exp\left[-\frac{1}{8}\xi r^2\right] \; ; \; \xi \simeq 1.69 \cdot 10^{-40} \frac{\kappa^2 n_0}{T} \; (m^{-2}) \; . \quad (17)$$

The numerical calculations of Eq. (15) show that the approximation (17) is valid near the stellar plasma center (for $x < 4$ ). The numerical calculations for $x \geq 8$ were found to be equal approximately to the solutions by Eq. (16). We can combine these asymptotic solutions into an approximate analytical result for all the range of $x$ values as:

$$\theta(x) = \frac{n_i(x)}{n_0} = Exp\left[-\frac{1}{8}x^2\right](for \;\; 0 \leq x \leq \infty) + \frac{2}{x^2}(for \;\; 4 \leq x \leq \infty) \quad . \quad (18)$$

By comparing the values calculated by Eq. (18) with the accurate numerical values (obtained by using Eq. (15)) we find that the analytical equation (18) gives fair good results (within accuracy of few percent) for the whole range of $x$ values

The use of the function: $\theta(x) = \frac{n_e(x)}{n_0}$ is remarkable. It gives also the plasma mass density $\rho(kg \cdot m^{-3}) = n_e \kappa m_n$ as function of the distance $r(m) = \frac{x}{\sqrt{\xi}}$ from the stellar plasma center which is proportional to the electron density $n_0(m^{-3})$ at its center. One should notice that the density of electrons $n_i(x)$ for $x \gg 1$ is decreasing inversely proportional to $x^2$ .

We calculated the mass of the stellar plasma separately for its central part and for its long tail.

**The mass of the central part of the low-density star plasma**

Using Eq. (17) the total number of electrons $N_{e,cent}$ in the central part of the present plasma is calculated by inserting the value of $\xi$ , given by Eq. (13) and performing the integral:

$$N_{e,cent} = n_0 \int_0^\infty Exp\left[-\frac{1}{8}\xi r^2\right] 4\pi r^2 dr = n_0 \frac{\pi\sqrt{\pi}}{(1/8)^{3/2} \xi^{3/2}} \simeq 5.73 \cdot 10^{61} \left(\frac{T}{n_0}\right)^{3/2} \cdot \frac{n_0}{\kappa^3}$$

By multiplying by $\kappa m_N$ we get for the central part mass $M_{cent}$ of this star

$$M_{cent}(kg) \simeq 4.8 \cdot 10^{34} \left(\frac{T}{n_0}\right)^{3/2} \cdot \frac{n_0}{\kappa^2} \quad . \quad (19)$$

For completely ionized plasma for which $\kappa = 1$, temperature $T = 6 \cdot 10^4 (K)$ and ionized electrons density $n_0 = 10^6 (m^{-3})$ we get $M_{cent} = 1.43 \cdot 10^{33} \cdot 10^6 (kg)$ which is approximately $715 \cdot 10^6$ times the sun mass. This result can be changed according to the parameters $T$ and $n_0$ of Eq. (19) but we find that the mass in the central part of the stellar low density plasma profile is quite large. In the next calculation we find that the mass in the long stellar plasma tail is also quite large.

**The mass of the tail of the low-density star plasma**

According to Eq. (16), the number of electrons in the tail of the low-density star plasma $N_{e,tail}$ is given by

$$N_{e,R} = n_0 \int_{R_{min}}^{R} \frac{2}{\xi r^2} 4\pi r^2 dr = \frac{8\pi n_0 (R - R_{min})}{\xi} \quad ; R = \frac{x}{\sqrt{\xi}} ; R_{min} = \frac{x_{min}}{\sqrt{\xi}} \quad . \quad (20\text{-a})$$

$$N_{e,tail} = n_0 \int_{R_{min}}^{R_{max}} \frac{2}{\xi r^2} 4\pi r^2 dr = \frac{8\pi n_0 (R_{max} - R_{min})}{\xi} \quad ; R_{max} = \frac{x_{max}}{\sqrt{\xi}} \quad . \quad (20\text{-b})$$

Here $N_{e,R}$ is the number of ionized electrons in the stellar plasma tail within the distance $R \gg R_{mn}$, $N_{e,tail}$ is given by the total number of ionized electrons in the dark halo with radius $R_{max} = \frac{x_{max}}{\sqrt{\xi}}$. The value of $R_{min}$ can be estimated from Eq. (17) where $R_{min} = \frac{x_{min}}{\sqrt{\xi}}$ ; $x_{min} \simeq 4$. We find here the interesting point that the amount of electrons for spherical shell with thickness $dr$ is constant, so that the divergence of the integral is prevented by maximal value $R_{max} = \frac{x_{max}}{\sqrt{\xi}}$. This maximal value is assumed at the present discussion as an empirical constant equal to the radius of the dark halo. It might be related to truncation of the Boltzmann distribution given by $x_{mx}$ to be discussed in one of the the following sections. Assuming extremely long distances $R$ and $R_{max}$ for which the lower limit in Eqs. (20-a) and (20-b) can respectively, be neglected we get for the mass of the stellar plasma tail the relations

$$M_R = \frac{8\pi n_0 R \kappa m_N}{\xi} \quad . \quad (21\text{-a})$$

$$M_{tail} = \frac{8\pi n_0 R_{max} m_N \kappa}{\xi} = \frac{8\pi n_0 x_{max} m_N \kappa}{\xi^{3/2}} \simeq 1.91\kappa^2 \cdot 10^{34} \frac{T^{3/2}}{\sqrt{n_0}} x_{max} (kg) \quad . \tag{21-b}$$

Here $M_R$ is the mass of the stellar plasma tail within a distance $R \gg R_{min}$. $M_{tail}$ is the total mass of the stellar mass tail where $R_{max} = \frac{x_{nax}}{\sqrt{\xi}}$ is the radius of the stellar plasma tail representing dark halo.

One should notice that the mass in the tail of the low density stellar plasma profile is proportional to $\frac{T^{3/2}}{\sqrt{n_0}}$ and is quite large for relatively large temperatures and low plasma density! Under the conditions $T = 6 \cdot 10^4 (K)$; $n_0 = 10^6 (m^3)$; $x_{max} \simeq 100$; $\kappa = 1$ this value is equal to $1.4 \cdot 10^{10}$ times the sun mass. It may increase for larger values of $x_{max}$ and lower values of the ionized electrons density $n_0$ so that it can be in the order of galaxy mass and even larger.

## 3. RELATIONS BETWEEN DARK HALOES AND LOW DENSITY STELLAR PLASMA AND POSSIBLEE RELATIONS BETWEEN DARK HALOES AND GALAXIES STARS

Rotation curves (RC) describe the rotational velocities of objects in a galaxy as function of their radial distance from the galaxy center. Various methods for deriving RC were described in the extensive literature on this subject [21-25]. By studying many galaxies it was found that the RC velocities are nearly constant, or "flat" with increasing distance away from the galactic center. This result is counterintuitive since based on Newtonian motion the rotational velocities would decrease for distances far away from the galactic center. By this argument the flat rotation curves suggest that each galaxy is surrounded by large amount of dark matter. There are various models describing dark matter haloes [9-14]. In the previous work [1] I related the dark matter and the dark haloes to low density ionized transparent plasma. The absorption and emission of such plasma is proportional to multiplication of the electrons and ions extremely low densities so that such effects can be neglected. But, as such low density plasma extends over extremely large volumes gravitational effects become important.

Considering a small mass with mass $m(kg)$ rotating in circular motion with radius $R$ around a big spherical mass $M(kg)$ then the Kepler potential is given by $m\dfrac{GM_{R,stars}}{R}$ where $M_{R,stars}$ is the big spherical mass within a distance $R$. The rotational kinetic energy of the small mass is given by $\dfrac{1}{2}mv_{rot}^2$. Due to the virial theorem for Kepler motion the kinetic energy is equal one half of the potential energy so we get

$$\frac{1}{2}mv_{rot}^2 = \frac{1}{2}m\frac{GM_{R,stars}}{R} \rightarrow v_{rot}^2 = \frac{GM_{R,stars}}{R} \quad . \tag{22}$$

Eq. (22) gives the Kepler rotational velocity under the approximation of neglecting deviation from spherical symmetry. By adding to the galaxy stars mass $M_{R,stars}$ the dark matter halo $M_{R,halo}$ (described as dark matter) Eq. (22) should be exchanged into

$$\frac{1}{2}mv_{rot}^2 = \frac{1}{2}m\frac{GM_{R,stars}}{R} + \frac{1}{2}m\frac{GM_{R,halo}}{R} \rightarrow v_{rot}^2 = \frac{GM_{R,stars} + GM_{R,halo}}{R} \quad . \tag{23}$$

Since the galaxy stars have usually a disk shape the use of Eq. (22) can be considered only as rough approximation. On the other hand we estimate that the dark halo has spherical symmetry as might follow from gravitational profiled related to dark haloes [10-14]. Due to the existence of dark matter the gravitational forces operating on the mass $m(kg)$ are composed of two parts: a) The Kepler attractive forces between the small mass and the galaxy high density stars. b) There is an additional attractive force between the small mass and the invisible dark matter mass (according to the present analysis low density plasma) included within a distance R. While the idea that RC velocity measurements are affected by such dark matter is quite common in the published literature [21-25], the physical nature of such matter is not clear and is, today, under many debates. One should take into account that historically the behavior of RC measurements which is in contradiction with Kepler motion was the first reason for introducing the dark matter which was considered as the missing mass (see e.g. [26]).

The physics of the high density stars is well understood. Energy generation is in the hearth of stars. It provides the energy that we see as light and it also supplies the heat and pressure that supports stars structures. The power source for stars is the thermonuclear fusion which leads to extremely high temperature in the cores of such stars. On the other hand the source of energy in dark matter is not clear. Much effort is spent both theoretically and experimentally to find new kinds of non-baryonic interactions leading to dark matter, but there is not yet any evidence to the existence of such forces. According to the present approach there is not any separate source of energy to dark matter but it is obtained from the energy of the high density stars and from the way by which stars are created. In astronomy it has been claimed "that galaxy forms within a dark halo" [9]. The main parameter on which the low density plasma depends is temperature, considered in our isothermal free electrons model to be a certain constant for each galaxy. It seems that the temperature of the low density plasma (and correspondingly its Boltzmann energy) is related to the luminosity of the galaxy stars. In astronomy, luminosity is defined as the total amount energy emitted per unit of time by a star, galaxy, or other astronomical object. "Brightness in astronomy, refers to the object brightness i.e. how bright an object appears to an observer. Brightness depends on both the luminosity of the object and on the distance between the object and observer, and also on any absorption of light along the path from the object to observer. Hertzsprung-Russel (HR) diagrams plot luminosity in certain units against temperature [16, 17]. Usually such diagrams are made in the range of temperatures: $3000 \leq T \leq 40000\,(K)$. There is strong empirical evidence that RC velocities are larger for galaxies with larger luminosity or correspondingly with larger brightness or larger mass and vice versa (see e. g. the empirical CR velocities measured in [21-25] and the discussions about Tully-Fisher relations, e.g. in [27-28]). According to the present approach the temperature (and correspondingly the Boltzmann free electrons energy) of low density plasma is related to a certain energy balance between the low density plasma temperature and galaxy stars luminosity (or the mass [28]).

While the high density stars exist in relatively small parts of the galaxy the dark matter exists everywhere in the galaxy and also beyond it as dark halo. In the present analysis this dark matter is represented by very low density stellar plasma which is invisible but has large gravitational forces due to its extreme volume. In the present analysis it has been claimed that the total attractive force affecting the rotation curves measurements is produced by the superposition of the Kepler potential and that of the low density stellar plasma potential. We follow here the idea that the low density plasma behaves as "collision less" plasma where the "collisions" between the compact high density stars and the low density plasma can be neglected. Such relation appearing in astronomical scale of lengths is similar to that of the "collision less" plasma in laboratories on earth where the collisions of the plasma with atoms or molecules can be neglected. It seems therefore that the separation of the potential of compact high density stars from that of the low density plasma is based on a common feature of plasma.

A general theory of RC velocities measurement should include the superposition of the gravitational forces produced by the high-density compact stars with that of the dark halo. In the present work I restrict the analysis to non-rotating dark halo. For haloes which are at extremely large distances from the galaxy center and which are beyond the compact stars the dominant effects will be mainly those of the dark haloes. We take into account also the conclusion in astronomy that the dark halo mass is equal approximately to 4 times the Kepler mass. We concentrate in the following analysis at the effects of the dark halos at very long radial distances at which the Kepler forces are quite small but they should be taken into account in more general analysis.

The present theory analyzed in section 2 for low density plasma has the advantage that it is developed from first physical principles while other descriptions of the halo density profiles [10-14] simulate gravitational observations without information about the composition of this matter. They seem to be variations of the present low-density profiles. Some mass density profiles [14] have exponential dependence similar to Eq. (17) (taking into account Eq. (4), where the mass density is proportional to the ionized electrons densities) but such profiles are proportional to $\exp[-Ar^\alpha]$ where $A$ and $\alpha$ are empirical constants. Some mass density profiles [12, 13] are similar to Eq. (16), (where the mass density profile is proportional to the ionized electrons density), but with high powers in the denominator. It is interesting to find that the mass density profile of the present work given by Eq. (16) is similar to that obtained by Burkert [12] which studied the structure of dark matter haloes in dwarf galaxies and arrived at isothermal density profile

$$\rho(r) = \frac{\rho_0}{1+\left(r^2/r_e^2\right)}, \qquad (24)$$

where $r_e$ is the core radius and $\rho_0$ is the density in the center of $\rho(r)$. This density profile is similar to the present low density profile as the tail of Eq. (16) proportional to $1/x^2$ can be related to Eq, (24) by the equality $1/x^2 = 1/\left(r\sqrt{\xi}\right)^2 = \frac{1/\left(\sqrt{\xi}\right)^2}{r^2} = \frac{(r_0)^2}{r^2}$ where $1/\sqrt{\xi} = r_0$ represents the central stellar plasma radius. Also, in the center of low-density stellar plasma we have density $n_0 \kappa m_N$ which is given in Eq. (24) by $\rho_0$ but the present density profile has a certain exponential dependence near the star plasma center (see Eq. (17)). Burkert in the abstract of his article [12] claimed: "that dark matter must be baryonic." We should take into account that for the total mass $M_{halo}$ of the halo of Eq. (24) we need to use the integral $M_{halo} = 4\pi \int_0^\infty r^2 \rho(r) dr = \int_0^\infty \frac{4\pi r^2 \rho_0}{1+(r^2/r_e^2)} dr$ and this integral will slowly diverge for $r \to \infty$. Burkert [12] and others [13] eliminated this divergence by adding high powers of $r$ in the denominator of Eq. (24). This divergence is prevented in the present model of low density plasma by the truncation of the Boltzmann distribution analyzed as follows.

**Truncation of the Boltzmann distribution**

Eq. (11) can be transformed to

$$\ln\left(\frac{n_0}{n_e(x)}\right) = \left(\frac{\kappa m_N}{k_B T}\right) \phi(x) = b\phi(x) \quad . \tag{25}$$

Under the condition $x \gg 1$ by using Eq. (16), the equality $\left(\frac{n_0}{n_e(x)}\right) \simeq \frac{x^2}{2}$ can be inserted in Eq. (25) to give

$$\ln\left(\frac{n_0}{n_e(x)}\right) = \ln\left(\frac{x^2}{2}\right) \simeq 2\ln x - \ln 2 \simeq \left(\frac{m_N \kappa}{k_B T}\right) \phi(x) \simeq b\phi(x) \quad . \tag{26}$$

We find that under the approximation $x \gg 1$, $2\ln x - \ln 2$ is equal to the potential $\phi(x)$ multiplied by the parameter $b = \left(\frac{\kappa m_N}{k_B T}\right)$ so that

$$\phi(x) = \left(\frac{k_B T}{\kappa m_N}\right)(2\ln x - \ln 2) \quad ; \quad for \quad x \gg 1 \quad . \tag{27}$$

Numerical -solutions of Eqs. (14-a) and (14-b), for the potential $\phi(x)$ at certain temperatures, were obtained by using Mathematica NDSolve and were found to give equivalent results to those of the asymptotic equation (27) for $x \gg 1$.

The potential-energy for one electron is given by multiplying Eq. (27) by $\kappa m_N$. Then, the potential for one electron is given as

$$\phi_e(x) = k_B T \left( 2\ln x - \ln 2 \right) (J) \quad ; for\ x \gg 1 \quad . \tag{28}$$

The maximal value $x_{max}$ for which Eq. (28) is valid can be obtained by the virial theorem by which the maximal electron kinetic energy (corresponding to the maximal velocity $v_{max}$ in Eq. (1) is equal to one half of the potential of Eq. (28). Then, we get for Hydrogen plasma (Z=1):

$$k_B T \left( 2\ln x_{max} - \ln 2 \right) \simeq \left( 2 \cdot 10^6 \right)^2 m_e (J) \to 2\ln x_{max} - \ln 2 = \frac{2.64 \cdot 10^5}{T} \quad . \tag{29}$$

For $T = 20000(K)$ we get $x_{max} \simeq 735$ while for $T = 60000(K)$ we get $x_{max} \simeq 9.71$. We find that $x_{max}$ has a strong dependence on temperature. The present model of free electrons breaks down approximately for $2\ln x_{max} - \ln 2 \simeq 1 \to T_{max} \simeq 2 \cdot 10^5 (K)$ but already for $T \geq 60000(K)$ the effects of Boltzmann distribution truncation are quite large. The present truncation of the Boltzmann distribution represents escaping of ionized electrons with very high velocities from the stellar plasma. Such effect occurs in evaporative cooling [29] and here we have a similar effect.

In the previous work [1] we analyzed, for low density ionized plasma, stimulated bremsstrahlung absorption and emission of radiation processes under Boltzmann statistics. Since these effects are proportional to the ionized electrons density squared (i.e. for multiplication of the ionized electrons densities by the ions densities) they can be neglected for extremely small ionized electrons densities. As our analysis is based on free electrons model we would like to apply the free electrons optical model [18] in the next section for treating further the absorption of radiation for low density plasma according to this model.

## 4. TRANSPARENCY OF LOW DENSITY STELLAR PLASMA AND EM WAVE DISPERSION IN RADIO FREQUENCY REGION ABOVE AND NEAR PLASMA FREQUENCY

By using for the stellar-plasma the free electron model conditions and assuming damped oscillator for the dipoles medium, the dielectric constant $\varepsilon$ for monochromatic EM waves propagating in the plasma is given by [18]

$$\varepsilon = \left(1 - \frac{\omega_p^2}{\omega^2 + i\gamma\omega}\right) \quad ; \quad \omega_p^2 = \frac{n_e e^2}{\varepsilon_0 m_e} \cong 3.183 \cdot 10^3 n_e \tag{30}$$

Here, $\varepsilon$ is the complex dielectric constant, $\omega\,(rad/\sec)$ is the electromagnetic wave frequency, $\gamma\,(\sec^{-1})$ the decay constant, $e$ the electron charge, $\omega_p^2$ is the plasma frequency squared of the free electron ionized plasma and the subscript $p$ refers to the medium polarization. The real and imaginary parts of the complex dielectric constant $\varepsilon = \varepsilon_1 + i\varepsilon_2$ are given by

$$\varepsilon_1(\omega) = \left(1 - \frac{\omega_p^2}{\omega^2 + \gamma^2}\right) \quad ; \quad \varepsilon_2(\omega) = \frac{\omega_p^2 \gamma}{\omega(\omega^2 + \gamma^2)} \quad . \tag{31}$$

For low-density plasma, for which the following conditions are satisfied

$$\omega \gg \omega_p \simeq 56.42\sqrt{n_e}\ (\sec^{-1}) \quad ; \quad \gamma \ll \omega \quad , \tag{32}$$

we get: $\varepsilon_1(\omega) = 1$ and $\varepsilon_2(\omega) \to 0$ so that the plasma is transparent for most of the EM spectrum. For low density plasma for which the average density is $\bar{n}_e = 10^6\ (m^{-3})$ the plasma frequency and wavelength are given, respectively, by $\omega_p \simeq 5.642 \cdot 10^4\ (\sec^{-1})$ $\lambda_p \simeq 3.34 \cdot 10^4\ (m)$, but for dark haloes the electron density and the plasma frequency might be much smaller. For frequencies $\omega$ which are tending to $\omega_p$ and are in the far radio frequency region we might obtain EM dispersion effects which are analyzed as follows. We can use the relations

$$\sqrt{\varepsilon} = \sqrt{\varepsilon_1 + i\varepsilon_2} = \eta + i\kappa \quad . \tag{33}$$

Here $\eta$ is the index of refraction and $\kappa$ is the extinction coefficient. The index of refraction $\eta$ and the extinction coefficient $\kappa$ are related to $\varepsilon_1$ and $\varepsilon_2$ by [18]

$$\eta^2 = \frac{\varepsilon_1}{2} + \frac{1}{2}\sqrt{\varepsilon_1^2 + \varepsilon_2^2} \quad ; \quad \kappa = \frac{\varepsilon_2}{2\eta} \quad ; \quad \varepsilon_1 = \eta^2 - \kappa^2 \quad ; \quad \varepsilon_2 = 2\eta\kappa \quad . \tag{34}$$

The extinction coefficient $\kappa$ is linked to the absorption coefficient $\alpha$ of Beer's law (describing the exponential attenuation of the intensity of EM beam $I(d)$ propagating in a distance $d$ through the medium via the relation $I(d) = I_0 e^{-\alpha d}$) by [18]

$$\alpha(\omega) = \frac{2\kappa(\omega)\omega}{c} \rightarrow \kappa(\omega) = \frac{\alpha(\omega)c}{2\omega} = \frac{\alpha(f)c}{4\pi f} \quad ; \quad f = \omega/2\pi \quad . \tag{35}$$

Equations. (30-35) are used [18] for treating EM wave propagation under the free electrons model. Here we apply them for analyzing EM waves, propagation through low-density plasmas.

Thermal free-free bremsstrahlung power absorption (i.e. inverse bremsstrahlung absorption) per unit frequency $f$ ($f = 2\pi\omega$) and unit volume ($m^3$) is given by [30]

$$\frac{dW}{dtdVdf}(J/m^3) = 4\pi\alpha_f B_f(T) \quad , \tag{36}$$

where $\alpha_f$ is the free-free absorption coefficient and $B_f(T)$ is the mean energy density at temperature T given by Planck's law

$$B_f(T) = \frac{2hf^3/c^2}{\exp(hf/k_B T) - 1} \quad . \tag{37}$$

As we are interested in the low frequency region for which $hf/k_B T \ll 1$ we are in Rayleigh regime for which $(1 - e^{-hf/k_B T}) \approx \frac{hf}{k_B T}$ and under this condition we get ( in the CGS units) [30]

$$\alpha(f) = 0.018 \cdot T^{-3/2} Z^2 n_e n_i f^{-2} \bar{g}_f \simeq 0.018 \cdot T^{-3/2} (n_e)^2 f^{-2} \quad , \tag{38}$$

where for ionized low density Hydrogen plasma we assumed $Z = 1$, $n_e = n_i$, and $\bar{g}_f$ is nearly 1 (under the condition $\frac{e^2}{4\pi\varepsilon_0 \hbar v} < 1$). By using, Eqs. (35) and (38) one gets [30] :

$$\kappa = \frac{\alpha(f)c}{4\pi f} = \frac{1}{4\pi} \cdot 0.018 \cdot T^{-3/2} (n_e)^2 f^{-3} c \quad . \tag{39}$$

For the range of densities and temperatures treated in the present analysis we find that $\kappa \rightarrow 0$, so that according to Eqs. (34), and (31) $\varepsilon_2 \rightarrow 0$, and $\gamma \rightarrow 0$, respectively. So, the dielectric constant is real and for frequencies above and near the plasma frequency

$$\varepsilon_1(\omega) \simeq \left(1 - \frac{\omega_p^2}{\omega^2}\right) \quad ; \quad \varepsilon_2(\omega) \simeq 0 \quad . \tag{40}$$

Using Eqs. (40), and (33) for frequencies near the plasma frequency, we get

$$\eta \simeq \sqrt{1 - \frac{\omega_p^2}{\omega^2}} \quad . \tag{41}$$

Such relation leads to divergence of the EM waves in the plasma above and near to the plasma frequency (phase velocity of the EM wave is given by; $c/\eta > c$ ). For frequencies below the plasma frequency there is no transmissions. Such effects are well known in optics for free electron model [18] when damping effects can be neglected. We obtained such effect for low density plasmas in the radio frequency region and if it can be obtained by dark matter it will proove the relevance of the present analysis to dark matter.

## 5. SUMMARY AND CONCLUSION

In the present work we extended the analysis of low density spherical stellar plasma [1] by adding numerical calculations and found important asymptotic equations for the density of ionized electrons- $n_i(r)\,(m^{-3})$ , mass density $\rho(r)\,(kg/m^3)$ and potential $\phi(r)$ as function of the distance $r$ from the stellar plasma center. Using stability analysis with Boltzmann statistics the density of ionized electron $n_i(r)$ was found to have an exponential dependence on the gravitational potential $\phi(r)$ given by Eq. (11). Using Poisson gravitational equation the equation for the potential $\phi(r)$ was given in Eq. (12). Differential equation was developed for the potential $\phi(x)$ in which we exchanged the distance $r$ into a pure number parameter $x$ and where $x = r\sqrt{\xi}$ , and $\xi$ is an extremely small number given by Eq. (13). Numerical calculations for the potential $\phi(x)$ were obtained as function of the parameter $\frac{\kappa m_N}{k_b T}$ where $m_N \cong 1.67 \cdot 10^{-27}\,kg$ is given by the nucleon mass, $T$ the absolute temperature, $k_B$ the Boltzmann constant and $\kappa$ is the average number of nucleons for one ionized electron as described in Eq. (4). Differential equation was obtained for the ionized electrons density $n_i(x)$ as function $x = r\sqrt{\xi}$ and the density of ionized electrons $n_0$ in the stellar plasma center taken as empirical parameter. The mass of the central mass profile was calculated by the use of the exponential equation (17) and the mass of the long stellar plasma tail was obtained by the dependence on $r$ given by Eq. (16). We found the asymptotic solution $n_i(x) = \dfrac{2}{x^2}$ for $x \gg 1$. The divergence obtained for the stellar plasma tail slow dependence on $r$ was eliminated by truncation of the Boltzmann distribution.

In astronomy it has been shown that most of the matter in the universe is in the form of dark matter which does not interact with the EM radiation. The physical composition of dark matter is not clear and there are many debates about it. In the present work it has been shown that low density plasma can enter into semi-equilibrium Boltzmann state (within astronomical time scales) and has the properties of dark matter. A general theory should include the superposition of the gravitational potentials introduced by the high density compact stars with those of the low density stellar plasma potentials. Such theories might be very complicated but for haloes which are at extremely large distance from the galaxy center the dominant effects will be those of dark matter and such dark haloes were analyzed in the present work by relating them to low density stellar plasma. Certain relations between the dark haloes and the luminosities of the high density stars were discussed. By using free electrons EM theory [18] it was shown in the present article that the low-density plasma is transparent in most of the EM spectrum. But near the low-density plasma frequency which is in the far radio frequency region dispersion effects related to the use of the approximation of Eq. (41) might be observed. Such observations if it can be obtained in dark matter it can give crucial evidence to the relevance of the low-density plasma to dark matter.

## REFERENCES

[1] Ben-Aryeh, Y. Transparency and stability of low density-plasma related to Boltzmann statistics and to thermal bremsstrahlung. *Scientific Reports, Nature,* **9**, 20384 (2019).

[2] W .J. Maciel. *Introduction to stellar structure* (Springer, Berlin, 2016).

[3] V. P. Krainov. Inverse stimulated bremsstrahlung of slow electrons under Coulomb scattering. J. Phys. B: At. Mol. Opt. Phys. **33**, 1585-1595 (2000)

[4] V.B. Berestetskii, A.M. Lifschitz, and L.P. Pitaevskii. *Quantum Electrodynamics* (Oxford, Oxford, 1998),

[5] N.L. Manakov, A. A. Krylovetsky, and S. L. Marmo. Bremsstrahlung radiation from slow electrons in Coulomb field: Classical limitand quantum correction.J. Exp. Theor. Phys. **121**, 727- 736 (2015).

[6] P. J. Brussaard and Van De Hulst. Approximation formulas for non relativistic bremsstrahlung and average Gaunt factors for a Maxwellian electron gas. Reviews of Modern Physics **34,** 507-520 (1962).

[7] F.F. Chen. *Introduction to plasma physics and controlled fusion.* (Plenum, New York, 1984).

[8] A. A. Fridman. *Plasma chemistry.* (Cambridge University Press, Cambridge, 2008).

[9] R. H. Wechsler and J. L. Tinker. The connection between galaxies and their dark matter Halos. Annu. Rev. Astron. Astrophys.. **56**: 435-487 (2018).

[10] D. Merritt, A. W. Graham, B. Moore, J. Diemand, and B. Terzic, Empirical models for dark matter Halos. I. Non parametric construction of density profiles and comparison with parametric model The Astronomical Journal **132**: 2685-2700 (2006).

[11] A. W. Graham, D. Merritt, B. Moore, J. Diemand, and B. Terzic. Empirical models for dark matter Halos. II. Inner profiles slopes, Dynamical profiles, and $\rho/\sigma^3$ . The Astronomical Journal **132**: 2701-2710 (2006).

[12] A. Burkert, The structure of dark matter Halos in dwarf galaxies. The Astronomical Journal, **447:** L. 25-L.28 (1995).

[13] J. F. Navarro, C. S. Frenk, and S. D. M. White . A universal density profile from hierarchical clustering. The Astrophysical Journal **490**: 493-508 (1997).

[14] E. Retana-Montenegro, E. Van-Hese, G. Gentile, M. Baes, and F. Frutos- Alfaro, Analytical properties of Einasto dark matter haloes. A&A **540**, A70 (2012).

[15] C. O. Wright and T. G. Brainerd. Gravitational lensing by NFW Haloes. Astrophysicsl Journal **534**, 34-40 (2000).

[16] N. Langer, and R. P. Kudritzki. The spectroscopic Hertzsprung –Russel diagram. Astronomy & Astrophysics **564**, A52 (2014).

[17] P. Lena, *Observational Astrophysics* (Cambridge University Press, Cambridge , 1995).

[18] S. A. Maier. Plasmonics: Fundamentals and applications (Springer, Berlin, 2007).

[19] X. Er, and M. Rogers.Two families of astrophysical diverging lens models. *MNRAS* **475,** 867-878 (2018).

[20] X. Er, and X. M. Rogers. Two families of elliptical plasma lenses *MNRAS* **488**, 5651-5664 (2019).

[21] V. C. Rubin, N. Thonnard, and K. W. Ford. Rotational properties of 21 SC galaxies with a large range of luminosities and radii from NGC 4605 (R=4 kpc) to UGC 2885 (R=122kpc). The Astrophysical Journal **238,** 471-487 (1980).

[22] Y. Sofue, Rotation and mass in the Milky way and spiral galaxies. Pub. Astro. Soc. Japan **69,** R1 (1-35).

[23] Y. Sofue and V. Rubin. Rotation curves of spiral galaxies. Annu.Rev.Astron. Astrophys. **39**:137-174 (2001).

[24] E. Battaner and E. Florido, The rotation curve of spiral galaxies and its cosmological implications. Fund. Cosmic. Phys, **21**, 1-154 (2000).

[25] W. J. G. De Blok, F. Walter, E. Brinks, C. Tachternach, S-H. Oh, and C. Kennicutt, Jr. High-resolution rotation curves and mass model from Things. The Astronomical Journal **136**: 2648-2719 (2008).

[26] F. Zwicky. On the masses of nebulae and cluster nebulae. The Astrophysical Journal **86**, 217-246 (1937).

[27] R. B. Tully and J. R. Fisher. A New method of determining distances to galaxies. Astron. Astrophys. **54,** 661-673 (1977).

[28] S. S. McGaugh, J.M. Schombert, G.D. Bothun, and W. J. G. Blok. The baryonic Fisher-Tully relation. The Astrophysical Journal, **533:** L99-L102 (2000).

[29] O. J. Luiten, M. W. Reynolds, and J. T. M. Walraven. Evaporative cooling of a trapped gas. *Phys. Rev. A.* **53**, 381-389 (1996).

[30] G. B. Rybicky, and A. P. Lightman . *Radiative processes in astrophysics* (Wiley, Weinheim, 2004).